\documentclass[onecollarge,natbib]{svjour2}
\bibpunct{[}{]}{;}{n}{}{,} 
\smartqed  
\usepackage{graphicx}
\usepackage{amssymb,amsmath,epsfig,bm,pifont}
\usepackage{color}

\journalname{Few-Body Systems}
\begin{document}
\title{Identifying the Universal Part of TMDs
\thanks{Presented by the first author at Light-Cone 2015,  
        September 21-25, 2015, Frascati, Italy.}
}


\author{F.F. Van der Veken \and
        N.G. Stefanis
        }


\institute{F.F. Van der Veken \at
           Departement Fysica,
           Universiteit Antwerpen,
           B-2020 Antwerpen, Belgium \\
           \email{frederik.van.der.veken@cern.ch}
           \\
           \emph{Present address: CERN, (Meyrin site),
           Building 9, Room 1-012,  
           CH-1211 Geneve 23, Switzerland}
           \and
           N.G. Stefanis \at
           Institut f\"{u}r Theoretische Physik II,
           Ruhr-Universit\"{a}t Bochum,
           D-44780 Bochum, Germany \\
           \email{stefanis@tp2.ruhr-uni-bochum.de}
}

\date{Received: date / Accepted: date}

\maketitle

\begin{abstract}
\emph{We attempt to identify a path layout in the definition of
transverse-momentum-dependent T-odd parton distribution functions
(TMD)s which combines features of both, initial- and final-state
interactions, so that it remains universal despite the fact that
the Wilson lines entering such TMDs change their orientation.
The generic structure of the quark correlator for this path layout
is calculated.}

\end{abstract}

\section{Introduction}
\label{intro}
Collinear factorization is a fully grown and well understood framework
to describe inclusive processes like deep inelastic scattering (DIS).
The hadron is portrayed by a parton distribution function (PDF) which
only depends on the longitudinal momentum fraction $x$ shared by the
struck quark, as the transversal momentum components are integrated out.
Factorization is then achieved by convoluting the PDF with the hard
part of the interaction and integrating over $x$,
see \cite{Brambilla:2014jmp} for a recent review.

The collinear framework is however less suitable to describe high-energy
particle collisions that are not fully inclusive.
This is, because in this case one has to take into account the transversal
part of the quark momentum as well.
This enlarged framework is based on $\bm{k_{\perp}}$ -factorization and
employs PDFs which depend on the transverse momentum --- thus
transverse-momentum-dependent PDFs or TMDs for short
(see \cite{Boer:2011fh} for an exhaustive review).
Exclusive processes are extremely hard to treat because one deals with
the full nonperturbative quark-gluon dynamics which cannot be computed
from first principles and has to be modeled.
However, for some processes which are semi-inclusive and the set of the
final states is not fully integrated out but is neither fully identified,
factorization theorems have been proven.
Among them, the most well known example is semi-inclusive deep inelastic
scattering (SIDIS), which has been thoroughly treated in
\cite{Mulders:1995dh,Boer:1997nt,Boer:1999uu}.
However, a few years later it was realized due to the non-vanishing
Sivers effect \cite{Brodsky:2002cx} that the presence of the Wilson line
inside the TMD definition should be taken into consideration
\cite{Collins:2002kn}.
The reason is that the Wilson line entails the existence of T-odd TMDs,
breaking the universality of the TMD.
Indeed, while in SIDIS the Wilson line is future-pointing, in initial-state
processes like the Drell-Yan (DY), where a produced virtual photon decays
into a muon pair, the Wilson line is past-pointing, leading to an overall
sign change for T-odd TMDs \cite{Ji:2002aa,Belitsky:2002sm,Boer:2003cm}.
This came as a surprise because the universality of the TMD was presumed
taking recourse to the stringent validity of collinear factorization in DIS.

This important finding notwithstanding, it appears attractive to ask whether
it is possible to create a path layout that combines features of both initial
and final state interactions.
Such a geometrical structure of Wilson lines could eventually be associated
with a common and hence universal part of the SIDIS and the DY processes.
This paper is devoted to the conceptual investigation of this issue.
The treatment of the renormalization issues of the correlators, in
particular questions related to the geometrical features of the Wilson
lines --- the appearance of cusps and their anomalous
dimensions --- is relegated to the literature (see, for instance,
\cite{Boer:2011fh,Collins:2011,Cherednikov:2008ua,Stefanis:2012if}
and references cited therein).
A more general assessment and a fuller treatment of the
present approach will be given elsewhere.
The work is organized as follows.
In the next section, we recall the structure of the gauge-invariant quark
correlator.
The new path layout is discussed in Sec.\ \ref{sec:new-path}, whereas the
generic quark correlator employing the new path layout is considered in
Sec.\ \ref{sec:cor-new-path}.
Finally, Sec.\ \ref{sec:concl} contains our main conclusions.

\section{Gauge-invariant quark correlator}
\label{sec:quark-cor}
The key quantities to describe the three-dimensional dynamics of partons
within hadrons, e.g., a proton, in high-energy collisions is provided
by TMD distribution and fragmentation functions.
Typically, these quantities depend on a large scale, say, $Q^2$, which
can be the invariant mass in a Drell-Yan process, and
$q_{\perp}^{2}\ll Q^2$,
the transverse momentum of the lepton pair in the same process (with
analogous definitions for SIDIS).
The generic form of the TMD quark correlator is given by
\begin{eqnarray}
  \Phi_{ij}^{q[C]}(x, {\bf{k}}_{\perp};n)
\sim
  \int_{}^{}\frac{dz^- d^2 {\bf{z}}_{\perp}}{(2\pi)^3}
  e^{-ik^{+}z^{-} + i\bm{k}_{\perp}\cdot\bm{z}_{\perp}}
      \left\langle
      p| \bar{\psi}_{j}(z)\mathcal{U}(0\rightarrow z|C)\psi_{i}(0)
      |p\right\rangle_{z^+=0} \, ,
\label{eq:TMD-definition}
\end{eqnarray}
where the Wilson line to ensure gauge invariance is a path-ordered
exponential of the form
\begin{equation}
  \mathcal{U}(a\to b|C)
  =
  {\cal P} \exp \left[
                      ig \int_{a}^{b} dz \cdot A(z)
                \right]
  \label{eq:Wilson-line}
\end{equation}
and the path $C$ has to be taken along the color flow from $a$ to $b$
for each considered process.

\begin{figure}[t]
\centering
\includegraphics[width=0.425\textwidth]{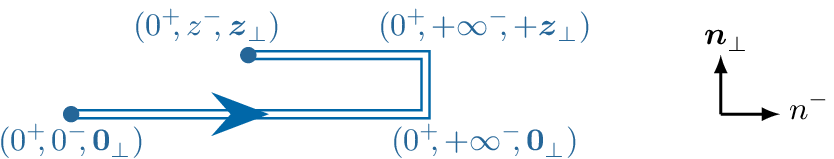}  
\hfill
\includegraphics[width=0.425\textwidth]{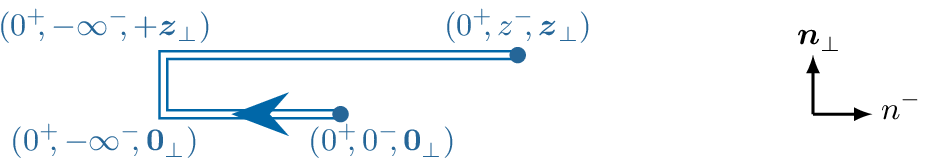}  
\caption{Path layouts pertaining to SIDIS (left) and the Drell-Yan
process (right)}
\label{fig:SIDIS-DY}
\end{figure}

As already mentioned, for processes with final-state interactions
(SIDIS) the path wraps around at $+\infty$ (left panel of
Fig.\ \ref{fig:SIDIS-DY}), whereas for processes with initial-state
interactions (DY process) --- right panel of Fig.\ \ref{fig:SIDIS-DY}
--- the path wraps around $-\infty$.
It can be shown
\cite{VanderVeken:2015cta,VanderVeken:2014eda,VanderVeken:2014kna}
that changing the endpoint of a linear Wilson-line segment from
$+\infty$ to $-\infty$ (or vice versa) amounts to the substitution
$n\to -n$ (where $n$ is the directional vector of the segment), i.e.,
$\mathcal{U}(a\to +\infty | n) = \mathcal{U}(a\to -\infty | -n)$.
Because the Wilson line is the only factor 
with an $n$-dependence, this relation is repeated in the correlator:
\begin{equation}
\label{eq: correlatorrelation}
  \Phi_{ij}^{q[C]}(x, {\bf{k}}_{\perp};n) \Big|_\text{DY}
=
  \Phi_{ij}^{q[C]}(x, {\bf{k}}_{\perp};-n) \Big|_\text{SIDIS}\, .
\end{equation}
The fully unintegrated quark correlator can be parameterized by a set
of scalar functions that are, after the $k^-$-integration,
related to their TMD counterparts \cite{Goeke:2005hb}.
The Lorentz structure of the correlator will only depend on contractions
of the vectors $p^\mu$, $k^\mu$, $S^\mu$ and $n^\mu$ with themselves
and with the Dirac basis.
In the case of an unpolarized proton the fully unintegrated
correlator can be cast in the form
\begin{align}
  \Phi(p,k,n) &= \nonumber
    M {A_1}
    + \gamma_\mu \, \left(
        p^\mu {A_2} + k^\mu {A_3}
        + \frac{M^2}{p\!\cdot\! n} n^\mu {B_1}
      \right)
  +
    \gamma_\mu \gamma_5 \,\frac{1}{p\!\cdot\! n}\varepsilon^{\mu\nu\rho\sigma}
      n_\nu p_\rho k_\sigma {B_4}
    \\
  &\qquad\qquad+
    i\gamma_{\mu\nu} \, \left[
      \frac{1}{M} p^\mu k^\nu {A_4}
      +\frac{M}{p\!\cdot\! n} \left(
        p^\mu {B_2} + k^\mu {B_3}
      \right) n^\nu
    \right]
    \, .
\end{align}
The important thing to notice here is that the directional vector
$n_\mu$ appears \textit{scaleless}, i.e., only via factors of
the form $\frac{n^\mu}{p\cdot n}$, being straightforward
to show that this remains true in the polarized case as well.
As the $k^-$-integration does not influence $n^\mu$, it is
tempting to conclude that the TMD correlator is invariant under a sign
change in $n$, and hence \emph{universal} by virtue of
Eq.\ \eqref{eq: correlatorrelation}.
However, one should not forget that the scalar functions themselves can
still depend on $n^\mu$.
Moreover, it has been demonstrated that those functions that are odd
under time reversal flip sign when interchanging past-pointing with
future pointing Wilson lines \cite{Collins:2002kn}.
This can be summarized in terms of the following relations
\begin{equation}
  f_{\text{T-even}}^{\text{DY}} = f_{\text{T-even}}^{\text{SIDIS}} \, ,
  \qquad\qquad\qquad
  f_{\text{T-odd}}^{\text{DY}} = -f_{\text{T-odd}}^{\text{SIDIS}} \, .
\end{equation}

\section{New path layout}
\label{sec:new-path}
\begin{figure}[t]
\centering
\includegraphics[width=0.4\textwidth]{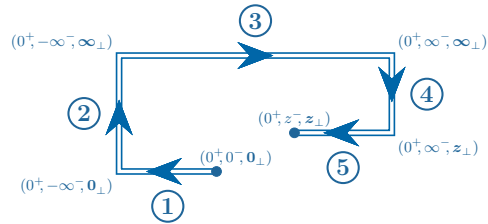}
\caption{
  The proposed new path layout for the Wilson line
}
\label{fig:new-layout}
\end{figure}

In what follows we will try to construct a path layout that encapsulates
features of both initial- and final-state interactions in a single frame.
First we note that due to the transitivity property of a Wilson line, the
layouts of SIDIS and DY can be easily related to one another
(see Fig. \ref{fig:SIDIS-DY-relation}):
\begin{figure}[t]
\centering
\includegraphics[width=0.65\textwidth]{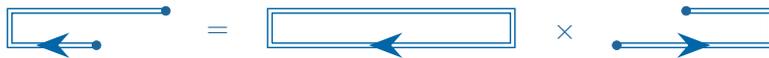}
\caption{
  Relation between the DY path layout (left-hand side) and
  that in SIDIS (right-hand side)
}
\label{fig:SIDIS-DY-relation}
\end{figure}
\begin{equation}
\label{eq:SIDIS-DY-relation}
  \mathcal{U}^\text{DY} = \mathcal{U}^\text{loop} \;
  \mathcal{U}^\text{SIDIS} \, ,
\end{equation}
where $\mathcal{U}^\text{loop}$ is a loop with infinite longitudinal
length (in the light-cone minus direction), and with
finite transversal width (equal to $\vec{z}_{\perp}$).
The overlapping parts cancel out on the right-hand side of Eq.
\eqref{eq:SIDIS-DY-relation}, resulting in the DY path layout.
However, this relation is not reproduced at the level of the correlator
$
 \Phi^\text{DY}
\neq
 \Phi^\text{loop} \;
 \Phi^\text{SIDIS}
$
because of the intricacies related to the splitting of expectation
values, so that in general
\begin{equation}
  \left\langle \mathcal{U}^\text{loop} \;
  \mathcal{U}^\text{SIDIS} \right\rangle
\neq
   \left\langle \mathcal{U}^\text{loop} \right\rangle
   \left\langle \mathcal{U}^\text{SIDIS} \right\rangle \, .
\end{equation}
We could at most hope that this relation holds up to power corrections
in
$\alpha_s$ and/or $N_{\!c}$.
Instead, we split the loop contour at transverse infinity into
parts, each in turn assigned to one of the two considered path layouts
so that one obtains the following configuration:
\begin{align}
  &\includegraphics[width=0.5\textwidth]{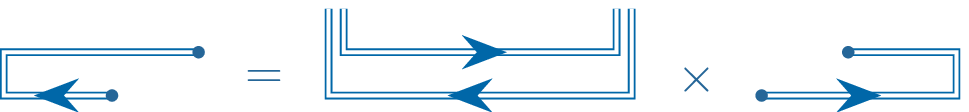} \\
  &\includegraphics[width=0.75\textwidth]{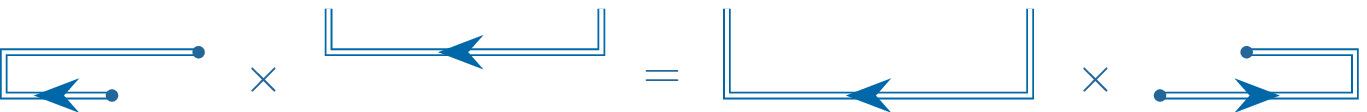}
\end{align}
The emerging layout represents a new contour pattern which can
be conceived of as being the ``golden mean'' between the SIDIS
and the DY path layouts.
Its structure is illustrated in Fig. \ref{fig:new-layout}: segments 1
and 2 correspond to the part on the left of the final-state cut
in the DY layout, while segments 4 and 5 refer to the
part on the right of the cut in the SIDIS layout.
Segment 3 vanishes in Lorentz and axial gauges, but is needed for gauge
invariance and in order to ensure that the full layout
reduces to the regular PDF layout after performing the
$\vec{k}_\perp$-integration (a simple straight line).
Let us emphasize that we do not pretend that the correlator built
from this layout should be considered as a physical quantity.
It is merely a mathematical construct out of academic interest, trying
to represent the core universal part of the regular quark correlator.
If we need to introduce a final-state cut, we draw it somewhere through
segment 3, the exact splitting point being irrelevant for practical
calculations.

\section{Generic quark correlator for the new path layout}
\label{sec:cor-new-path}
The next step is to retrieve the quark correlators for SIDIS and DY
from our new correlator.
We denote the latter, i.e., the correlator constructed from our new
path layout, as $\tilde{\Phi}$, so that we can write the
original correlators as a convolution between a (process-dependent)
correction factor and $\tilde{\Phi}$:
\begin{align}\label{eq:convolutions}
  \Phi^{\text{SIDIS}} &= C^\text{SIDIS} \otimes \tilde{\Phi} \, , &
  \Phi^{\text{DY}} &= C^\text{DY} \otimes \tilde{\Phi} \, ,
\end{align}
where the $C^i$ are expected to be perturbative quantities.
In leading order, all Wilson lines reduce to
unity and hence
$\Phi^{\text{SIDIS}} = \Phi^{\text{DY}} = \tilde{\Phi}$.
This implies that the correction factors at the leading-order level
are just delta functions:
\begin{equation}
  C^\text{SIDIS}_0
=
  C^\text{DY}_0
=
  \delta(1-x) \delta^{(2)}\!(\vec{k}_\perp) \, .
\end{equation}
A fruitful approach to continue to higher orders, is to focus on
parameters which are peculiar to each of the three correlators,
$\Phi^{\text{SIDIS}}, \Phi^{\text{DY}}, \tilde{\Phi}$.
In Feynman gauge (or another Lorentz, i.e., covariant gauge),
only segments 1 and 5 in Fig. \ref{fig:new-layout} survive.
What changes across different correlators is the orientation
of these segments, as depicted graphically in
Table \ref{tab:parameter-table}.
To maximize reproducibility, we assign different directional
vectors $n_1$ and $n_5$ to these segments.
This allows the retrieval of each correlator by a mere
substitution of $n_{1,5}$.
Similarly, in the light-cone gauge (or any other axial gauge),
only segments 2 and 4 contribute, and the differing parameter is the
transversal endpoint $r^-$.
Hence, we assign different endpoints $r_2$ and $r_4$ to these segments.

In what follows, we perform our investigation of the parametric
structure in Feynman gauge; the extension to the light-cone
gauge is, however, trivial.
Expanding Eqs. \eqref{eq:convolutions} in terms of powers of
$\alpha_s$, it is easy to show that the first-order correction
factors depend on the relevant substitution differences:
\begin{align}
  C^\text{SIDIS}_1 &\sim \Phi_1\big|_{n_1 \to -n_1} - \Phi_1 \, , &
  C^\text{DY}_1 &\sim \Phi_1\big|_{n_5 \to -n_5} - \Phi_1 \, ,
\end{align}
where $\Phi$ is the generic correlator, i.e., with both $n_1$ and $n_5$
left unspecified.
Now to continue, we note that the generic correlator
in leading order of $\alpha_s$ consists of three parts:
a term not depending on the directional vector,
and terms ``even'' or ``odd'' in $n$:\footnote{No other terms
appear at the first-order level due to
$(n_1)^2=(n_5)^2=n_1\!\cdot n_5 =0$.
One can keep the ``even/odd'' distinction at higher orders
using the \emph{sign} of $n_1$ and $n_5$ instead of
employing the vectors themselves.}
\begin{equation}
  \Phi
=
  \Phi_0 + \alpha_s \left(A + \frac{n_1+n_5}{2}B
  + \frac{n_1-n_5}{2}C\right)
  + \mathcal{O}(\alpha_s^2) \, ,
\end{equation}
where $A,B$, and $C$ are the same for all three correlators.
Comparing this generic form with Table \ref{tab:parameter-table}, we
observe that
\begin{subequations}
\begin{align}
  \tilde{\Phi}
& =
    \Phi_0 + \alpha_s \left(A - n^- B\right) + \mathcal{O}(\alpha_s^2) \, ,
&
  C^\text{SIDIS}_1
& =
    - \alpha_s n^- (B+C) \, ,
\\
  \Phi^{\text{SIDIS},\text{DY}}
& =
    \Phi_0 + \alpha_s \left(A \pm n^- C\right) + \mathcal{O}(\alpha_s^2) \, ,
&
  C^\text{DY}_1
& =
    - \alpha_s n^- (B-C) \, .
\end{align}
\end{subequations}
It is important to realize that $A$ is the only term that is
T-even, with both $B$ and $C$ being T-odd.
If $B$ would vanish, then $\tilde{\Phi}$ would be completely
universal.
A deep investigation of this term would help to shed light on
the universality properties of $\tilde{\Phi}$ and TMDs in more
general terms.
\begin{table}[t]
\centering
\includegraphics[width=0.55\textwidth]{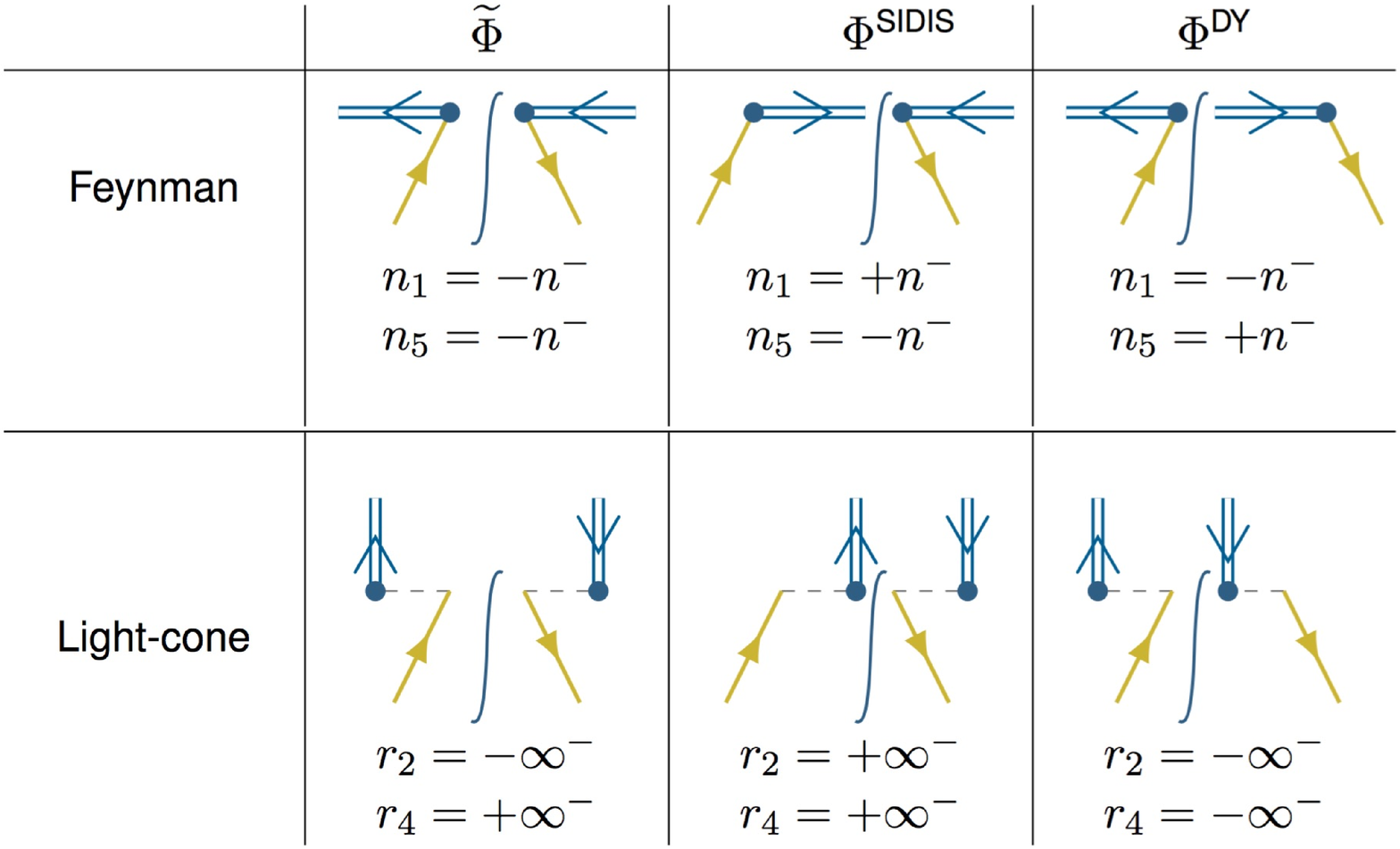}
\caption{
  Values of the parameters $n_1, r_2, r_4$ and $n_5$
  entering the various considered correlators and pertaining to
  two different gauges.
  The double lines denote Wilson lines
}
\label{tab:parameter-table}
\end{table}

\section{Conclusions}
\label{sec:concl}
We have constructed a new path layout, which combines features
of both initial- and final-state interactions entering
the calculation of quark correlators in SIDIS and DY-type processes
making use of Wilson lines to preserve gauge invariance.
The ensuing generic correlator provides a theoretical tool for the
investigation of the universality properties of the original correlators
considered in the SIDIS and Drell-Yan processes in connection with
particular process-dependent contours
\cite{Goeke:2005hb,Bacchetta:2004zf}.
The proposed path layout reveals universal features and enhances
the insight into the universality properties of the known correlators.
Adopting an approach which employs two parameters, we were able to
perform explicit calculations of these correlators in the Feynman
and also in the light-cone gauge and demonstrate the usefulness
of the proposed layout.
A more general investigation of the universal features of TMDs, which
includes the renormalization properties of the correlators, will be
given elsewhere.

\begin{acknowledgements}
F.\ F.\ Van der Veken gratefully acknowledges support from the
Gary T.\ McCartor Fund Fellowship and thanks the Research
Foundation-Flanders (FWO) for a travel grant.
He is also thankful to the members of the Institute for Theoretical
Physics II of Bochum University for their warm hospitality
during a research stay when most part of this investigation
was carried out.
\end{acknowledgements}

\end{document}